\documentclass[a4paper,11pt]{article}
\usepackage{pos}

\title{Discretization effects on nucleon root-mean-square radii
 from lattice QCD at the physical point}
\ShortTitle{Discretization effects on nucleon root-mean-square radii....}

\author*[\dag, a.b]{Ryutaro Tsuji}
\author[b]{Yasumichi Aoki}
\author[c]{Ken-Ichi Ishikawa}
\author[d]{Yoshinobu Kuramashi}
\author[a]{Shoichi Sasaki}
\author[d]{Eigo Shintani}
\author[e.d]{Takeshi Yamazaki}
\note[\dag]{Present adress is \textit{RIKEN Center for Computational Science, 650-0047, Kobe, Japan}.}
\affiliation[]{\normalsize{\bf \sffamily \hspace{50mm}(PACS Collaboration)}}

\affiliation[a]{Department of Physics, Tohoku University,\\
  980-8578,Sendai, Japan}

\affiliation[b]{RIKEN Center for Computational Science,\\
  650-0047, Kobe, Japan}

\affiliation[c]{Core of Research for the Energetic Universe,\\
    Graduate School of Advanced Science and Engineering, Hiroshima University\\
  739-8526, Higashi-Hiroshima, Japan}

\affiliation[d]{Center for Computational Sciences, University of Tsukuba,\\
  305-8577, Tsukuba, Japan}

\affiliation[e]{Institute of pure and Applied Sciences, University of Tsukuba,\\
  305-8571, Tsukuba, Japan}

\emailAdd{tsuji@nucl.phys.tohoku.ac.jp}

\abstract{
    We present results for the axial-vector coupling and root-mean-square (RMS) radii of the nucleon obtained from 2+1 flavor lattice QCD at the physical point with a large spatial extent of about 10 fm. 
    Our calculations are performed with the PACS10 gauge configurations generated by the PACS Collaboration with the six stout-smeared $O(a)$ improved Wilson-clover quark action and Iwasaki gauge action at $\beta$ = 1.82 and 2.00 corresponding to lattice spacings of 0.085 fm and 0.063 fm, respectively. 
    We first evaluate the value of the axial-vector coupling of the nucleon ($g_A$).
    In addition,
    the isovector electric, magnetic and axial radii and magnetic moment from the corresponding form factors are also determined.
    Combining the results at $\beta=1.82$ and $2.00$,
    we finally discuss the finite lattice spacing effect.
    It was found that the effect on $g_A$ is kept smaller than the statistical error of 2\% while the effect on the isovector radii was observed as a possible discretization error of about 10\%, regardless of the channel. 
}

\FullConference{The 40th International Symposium on Lattice Field Theory (Lattice 2023)\\
July 31st - August 4th, 2023\\
Fermi National Accelerator Laboratory\\}

\begin{document}
\maketitle

\section{Introduction}
\label{sec:introduction}
In the standard model of modern particle physics, protons and neutrons, known as nucleons, are composite particles of quarks and gluons, and the interaction among them is formulated as Quantum Chromodynamics (QCD).
This indicates that the nucleon has a non-trivial structure due to the complex dynamics of QCD.
One of the topics that have recently come into the spotlight is the ``size'' of the nucleon such as electric $(\langle r_E^2 \rangle)$, magnetic $(\langle r_M^2 \rangle)$ and axial $(\langle r_A^2 \rangle)$ radii, which can be extracted as the root-mean-square (RMS) radius from the corresponding form 
factors~\cite{Meyer:2022mix}.

Recently, much effort has been devoted to the high-precision determination of the nucleon RMS radius in each channel using lattice QCD calculations.
%The lattice QCD community has computed the size of the nucleon.
%Towards the high-precision determination by lattice QCD,
In general, the observables determined from lattice QCD simulations are subject to both statistical and systematic uncertainties. 
%As for {\color{red}the nucleon RMS radii},
As for the systematic uncertainties on the nucleon RMS radius, it is known that there are four major sources: 1) chiral extrapolation, 2) finite size effect, 3) finite lattice spacing effect and 4) excited-state contamination.
%Recent lattice QCD calculations have succeeded in reproducing the results being consistent with the experiments~\cite{Jang:2019vkm, Djukanovic:2023jag, , RQCD:2019jai, Park:2021ypf},
%by taking into account the detailed analyses for the systematic uncertainties.
Although great efforts has been made to improve our knowledge of the nucleon,
it is not accurate enough to solve problems (e.g. the proton radius puzzle~\cite{Pohl:2010zza}) or to provide a high-precision input for current neutrino oscillation experiments~\cite{Tomalak:2023pdi}.
The most recent lattice QCD simulations done by NME~\cite{Park:2021ypf}, PNDME~\cite{Jang:2023zts} and Mainz~\cite{Djukanovic:2022wru, Djukanovic:2023beb, Djukanovic:2023jag} groups have reproduced the nucleon electric RMS radius, magnetic moment, magnetic RMS radius and axial radius with statistical precision of about 5\%
by taking into account the detailed analyses for the systematic uncertainties.
However,
percent-level accuracy for the electromagnetic sector and a few percent-level calculations for the rest are highly desirable for solving puzzles and tensions~\cite{Meyer:2022muy}.

In this work,
we present the result of our recent study on the nucleon form factors.
Our previous work~\cite{Shintani:2018ozy} has dealt with most uncertainties, but discretization uncertainties have not yet been examined.
Therefore we calculate on the second PACS10 ensemble in order to study the discretization uncertainties of the nucleon form factors.
Combining our previous results obtained from  
the coarser lattice spacing~\cite{Shintani:2018ozy},
the finite lattice spacing effects are 
investigated by comparing the two results.

\section{Method}
\label{sec:method}
We calculate the electric and magnetic form factors, $G_E(q^2)$ and $G_M(q^2)$ and the axial form factor $F_A(q^2)$.
The first two are relevant for the electron-nucleon scattering experiment, while the latter is important input for the weak process associated with the neutrino-nucleus scattering.

We simply focus on the isovector quantities, 
where there is no disconnected contribution in the exact SU(2) isospin limit~\cite{Sasaki:2003jh}. 
The isovector electric and magnetic form factors are given by the combination of proton's and neutron's form factors,
\begin{align}
    G^{v}_l(q^2) = G^{p}_{l}(q^2)-G^{n}_{l}(q^2),
    \quad l=\{E,M\}.
\end{align}
which can be used for comparison with experiments without
evaluating the disconnected contributions. 
As for the axial form factor, the axial vector coupling, $g_A = F_A(q^2=0)$, is experimentally well determined as $g_A=1.2756(13)$~\cite{ParticleDataGroup:2022pth}. 
Therefore, we also calculate this particular quantity as a good reference for verifying the accuracy and reliability of our calculations.

In this study,
the exponentially smeared quark operator $q_S(t,\boldsymbol{x})=\sum_{\boldsymbol{y}}A\mathrm{e}^{-B|\boldsymbol{x}-\boldsymbol{y}|}q(t,\boldsymbol{y})$
with the Coulomb gauge fixing is used for the construction of the nucleon interpolating operator as well as a local quark operator $q(t,\boldsymbol{x})$.
The nucleon two-point function with nucleon interpolating operator located at smeared (\textit{S}) or local (\textit{L}) source $(t_{\rm src})$, and local sink $(t_{\rm sink})$ is constructed as
\begin{align}
    \label{eq:two_pt_func}
    C_{XS}(t_{\mathrm{sink}}-t_{\mathrm{src}}; \boldsymbol{p})
    =
    \frac{1}{4}\mathrm{Tr}
    \left\{
        \mathcal{P_{+}}\langle N_X(t_{\mathrm{sink}}; \boldsymbol{p})
        \overline{N}_S(t_{\mathrm{src}}; -\boldsymbol{p})  \rangle
    \right\}\ \mathrm{with}\ X=\{S,L\},
\end{align}
where the nucleon operator with a three-dimensional momentum $\boldsymbol{p}$ is given for the proton state by
\begin{align}
    \label{eq:interpolating_op}
N_L(t,\boldsymbol{p})&
    =\sum_{\boldsymbol{x}}e^{-i\boldsymbol{p}\cdot\boldsymbol{x}}\varepsilon_{abc}
    \left[
        u^{T}_{a}(t,\boldsymbol{x})C\gamma_5d_b(t,\boldsymbol{x})
    \right]
    u_c(t,\boldsymbol{x})
\end{align}
with the charge conjugation matrix, $C=\gamma_4\gamma_2$. The superscript $T$ denotes a transposition, while the indices $a$, $b$, $c$ and
$u$, $d$ label the color and the flavor, respectively.

To calculate the isovector nucleon form factors, 
we evaluate the nucleon three-point functions, which are constructed with the spatially smeared source and
sink operators of the nucleon as 
\begin{align}
    \label{eq:three_pt_func}
    C^{k}_{\mathcal{O}_\alpha}(t; \boldsymbol{p}^{\prime}, \boldsymbol{p})
    =
    \frac{1}{4} \mathrm{Tr}
    \left\{
        \mathcal{P}_{k}\langle N_S(t_{\mathrm{sink}}; \boldsymbol{p}^{\prime})
        J^{\mathcal{O}}_{\alpha}(t; \boldsymbol{q}=\boldsymbol{p}-\boldsymbol{p}^{\prime})
        \overline{N}_S(t_{\mathrm{src}}; -\boldsymbol{p})  \rangle
    \right\}
\end{align}
with a given isovector bilinear operator defined
as $J^{\mathcal{O}}_\alpha = \bar{u}\Gamma^\mathcal{O}_\alpha d$
with $\Gamma^\mathcal{O}_\alpha = \gamma_\alpha$ and $\gamma_\alpha\gamma_5$
for the vector $(V_\alpha)$ and axial-vector $(A_\alpha)$.
In the above equation, 
$\mathcal{P}_k$ denotes the projection operator to
extract the form factors.
For the unpolarized case ($k=t$), $\mathcal{P}_t\equiv\mathcal{P}_{+}\gamma_4$ is chosen,
while $\mathcal{P}_{5z}\equiv\mathcal{P}_{+}\gamma_5\gamma_z$ is chosen
for the polarized case in $z$ direction ($k=5z$).
In a conventional way to extract the form factors, we take an appropriate combination of two-point function~(\ref{eq:two_pt_func}) and three-point function~(\ref{eq:three_pt_func}), 

\begin{align}
\mathcal{R}_{\mathcal{O}_\alpha}^{k}\left(t; \boldsymbol{p}^{\prime}, \boldsymbol{p}\right) = \frac{C_{\mathcal{O}_\alpha}^{k}\left(t; \boldsymbol{p}^{\prime}, \boldsymbol{p}\right)}{C_{SS}\left(t_{\mathrm{sink}}-t_{\mathrm{src}}; \boldsymbol{p}^{\prime}\right)}
\sqrt{\frac{C_{L S}\left(t_{\mathrm{sink}}-t; \boldsymbol{p}\right) C_{SS}\left(t-t_{\mathrm{src}}; \boldsymbol{p}^{\prime}\right) C_{L S}\left(t_{\mathrm{sink}}-t_{\mathrm{src}}; \boldsymbol{p}^{\prime}\right)}{C_{L S}\left(t_{\mathrm{sink}}-t; \boldsymbol{p}^{\prime}\right) C_{S S}\left(t-t_{\mathrm{src}}; \boldsymbol{p}\right) C_{LS}\left(t_{\mathrm{sink}}-t_{\mathrm{src}}; \boldsymbol{p}\right)}},
\end{align}
which yields the following asymptotic values in the asymptotic region 
($t_{\mathrm{sep}}/a\gg (t-t_{\mathrm{src}})/a \gg 1$):
\begin{align}
    \label{eq:ge_def}
    \mathcal{R}^t_{V_4}(t;\boldsymbol{q})
    &=
    \frac{1}{Z_V}
    \sqrt{\frac{E_N+M_N}{2E_N}}{G}^{v}_E(q^2),\\
    \label{eq:gm_def}
    \mathcal{R}^{5z}_{V_i}(t;\boldsymbol{q})
    &=
    \frac{1}{Z_V}
    \frac{-i\varepsilon_{ij3}q_j}{\sqrt{2E_N(E_N+M_N)}}{G}^{v}_M(q^2),\\
    \label{eq:fa_def}
    \mathcal{R}^{5z}_{A_i}(t;\boldsymbol{q})
    &=
    \frac{1}{Z_A}
    \sqrt{\frac{E_N+M_N}{2E_N}}
    \left[
        {F}_A(q^2)\delta_{i3}-\frac{q_iq_3}{E_N+M_N}{F}_P(q^2)
    \right],
\end{align}
where the renormalization factors are defined through the renormalization of the quark currents
$J^{\mathcal{O}}_\alpha=Z_O\widetilde{J}^{\mathcal{O}}_\alpha$.
The renormalization factor $Z_V$ and $Z_A$ are determined by the Schr\"odinger functional method (see Appendix E in Ref.~\cite{Tsuji:2023llh}).
We evaluate the values of the target form factor in the standard plateau method.

The RMS radius of a given form factor $G_l(q^2)$ can be determined from the slope of $G_l(q^2)$ at $q^2=0$,
            
\begin{align}
    \langle r^2_{l} \rangle
    &=
    -\frac{6}{G_{l}(0)}
    \left.
        \frac{dG_{l}(q^2)}{d q^2}
    \right|_{q^{2}=0}.
\end{align}

\noindent
Here we use the notation of $G_A\equiv F_A$ for the axial-vector form factor.
Finally in this study,
z-expansion method, which in known as an model-independent $q^2$-parameterization,
is employed for studying the $q^2$-dependence of each form factor.
For a detail of the z-expansion method,
see Ref.~\cite{Hill:2010yb, Tsuji:2023llh} and all relevant references therein.
%In this study,
%the z-expansion method, which is known as an model-independent analysis, is employed~\cite{Bhattacharya:2011ah, T2K:2023smv, Simons:2022ltq}.
%In the $z$-expansion method, 
%the given form factor $G(q^2)$ is fitted by the following functional form
%%
%\begin{align}
%    \label{eq:z-expansion}
%    G(q^2)
%    &=
%    \sum_{k=0}^{k_{\mathrm{max}}}c_kz(q^2)^k=
%    {c_0+c_1z(q^2)+c_2z(q^2)^2+c_3z(q^2)^3}+\dots,
%\end{align}
%%
%where a new variable $z$ is defined by a 
%conformal mapping from $q^2$ as
%%
%\begin{align}
%    \label{eq:conformal-map}
%    z(q^2)
%    =\frac{\sqrt{t_{\rm cut}+q^2}-\sqrt{t_{\rm cut}}}
%    {\sqrt{t_{\rm cut}+q^2}+\sqrt{t_{\rm cut}}}
%\end{align}
%%
%with $t_{\rm cut}=4m_\pi^2$ for $G_E$ and $G_M$, or with $t_{\rm cut}=9m_\pi^2$ 
%for $F_A$.,
%In this study, $k_{\mathrm{max}}=3$ is adopted.

\section{Simulation details}
\label{sec:simulation_details}
We mainly use the PACS10 configurations generated by the PACS Collaboration with the six stout-smeared ${\mathscr{O}}(a)$ improved Wilson-clover quark action and Iwasaki gauge action at $\beta=1.82$ and $2.00$ corresponding to the lattice spacings of $0.085$ fm (coarse) and $0.063$ fm (fine), respectively~\cite{Shintani:2018ozy, Tsuji:2022ric, Tsuji:2023llh}. 
When we compute nucleon two-point and three-point functions, the all-mode-averaging (AMA) technique~\cite{Blum:2012uh, Shintani:2014vja, vonHippel:2016wid} is employed in order to reduce the statistical errors significantly without increasing
computational costs. 
The nucleon interpolating operators defined in Eq.~(\ref{eq:interpolating_op}) are exponentially smeared with $(A, B)=(1.2,0.16)$ for $128^4$ lattice ensemble and $(A,B)=(1.2,0.11)$ for $160^4$ lattice ensemble. As for the three-point functions, the sequential source method is employed and calculated with $t_{\rm sep}/a=\{10,12,14,16\}$ for $128^4$ lattice ensemble and $t_{\rm sep}/a=\{13,16,19\}$ for $160^4$ lattice ensemble.

%
% TABLE 1
%
\begin{table*}[ht]
\caption{
Parameters of the second PACS10 ensemble. See Refs.~\cite{Shintani:2018ozy, Tsuji:2022ric, Tsuji:2023llh} for further details.
\label{tab:simulation_details}}
\centering
\begin{tabular}{cccccccc}
\hline \hline
 $\beta$ &$L^3\times T$ & $a^{-1}$ [GeV] &  $\kappa_{ud}$ & $\kappa_{s}$ &$c_{\mathrm{SW}}$ &  $m_\pi$ [GeV] \\
          \hline
     2.00& $160^3\times 160$ &3.1108(70)& 0.125814 & 0.124925 & 1.02 & 0.138\\
     1.82& $128^3\times 128$ &2.3162(44)& 0.126177 & 0.124902 & 1.11 & 0.135\\
\hline \hline
\end{tabular}
\end{table*}

\section{Numerical results}
\label{sec:numerical_results}
Combining our two results from large volume simulations at the 
fine and coarse lattice spacings,
we can discuss the finite lattice spacing effects on 
 $g_A$, $\mu_v$, $\sqrt{\langle (r^v_E)^2 \rangle}$, $\sqrt{\langle (r^v_M)^2 \rangle}$ and$\sqrt{\langle (r^v_A)^2 \rangle}$.
Recall that the continuum limit results are not yet determined in our study,
we only evaluate the differences between two results from different lattice spacing as a possible
size of the lattice spacing effect.

\begin{figure*}
\centering
\includegraphics[width=0.32\textwidth,bb=0 0 792 612,clip]{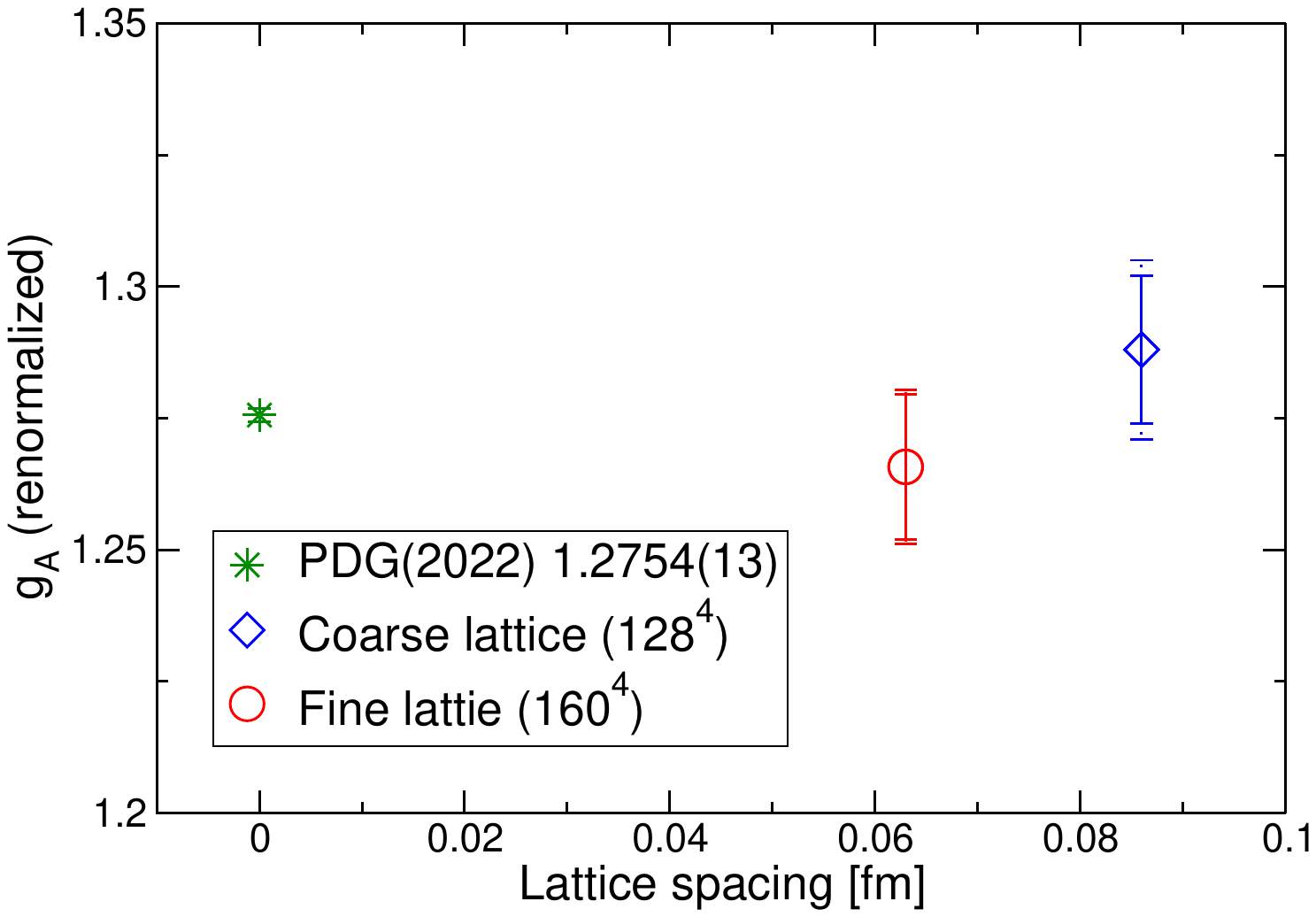}
\includegraphics[width=0.32\textwidth,bb=0 0 792 612,clip]{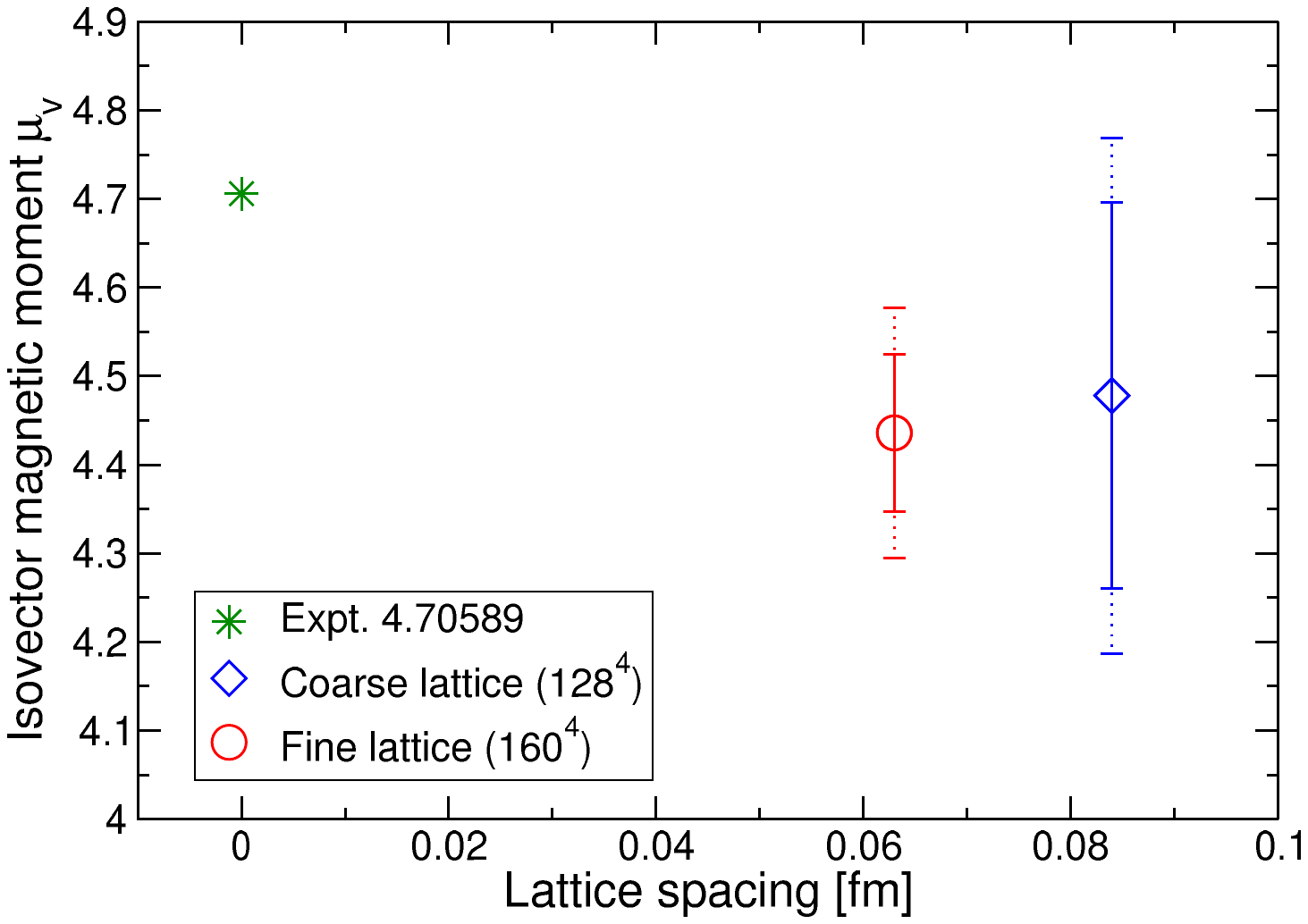}\\
\includegraphics[width=0.32\textwidth,bb=0 0 792 612,clip]{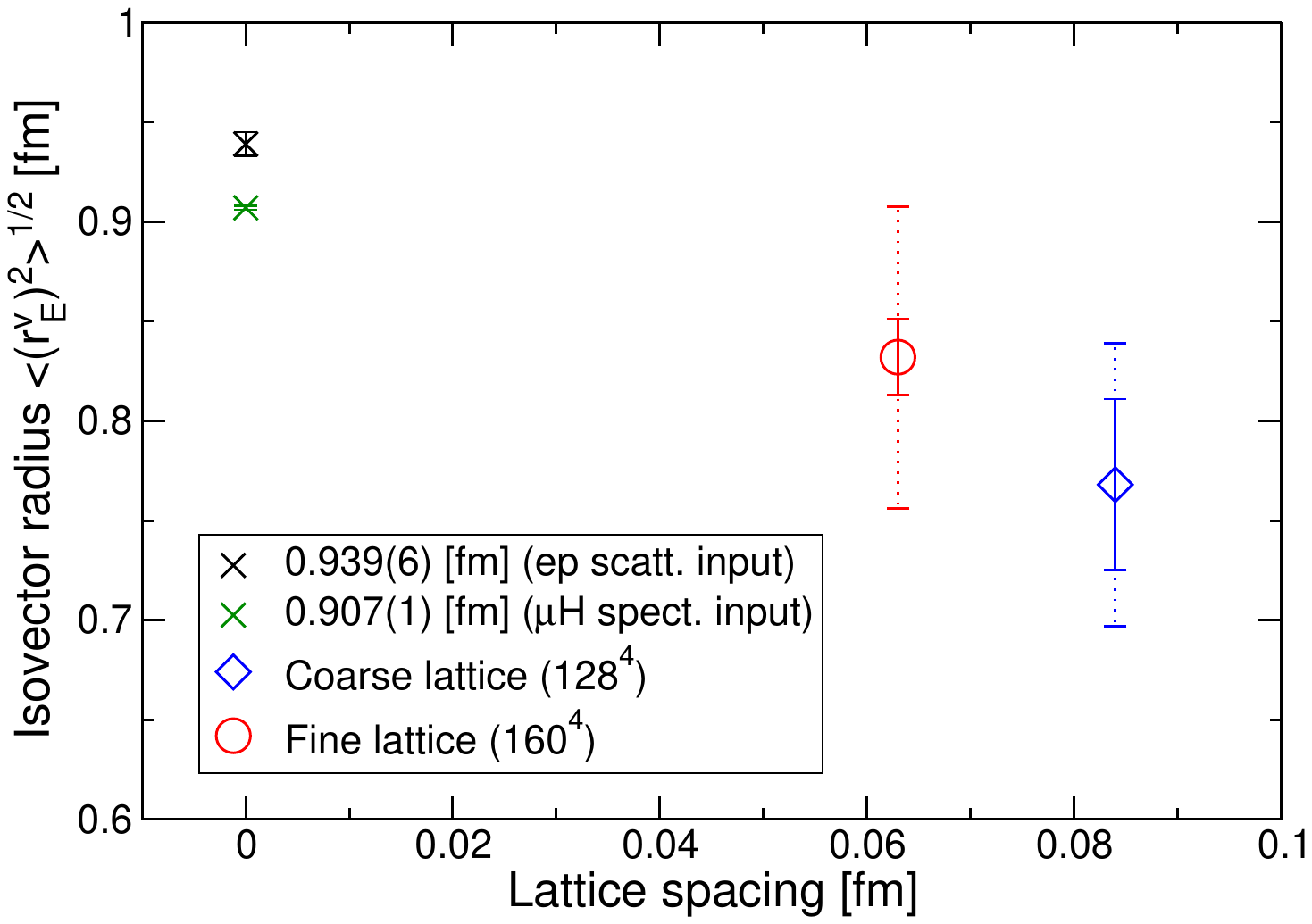}
\includegraphics[width=0.32\textwidth,bb=0 0 792 612,clip]{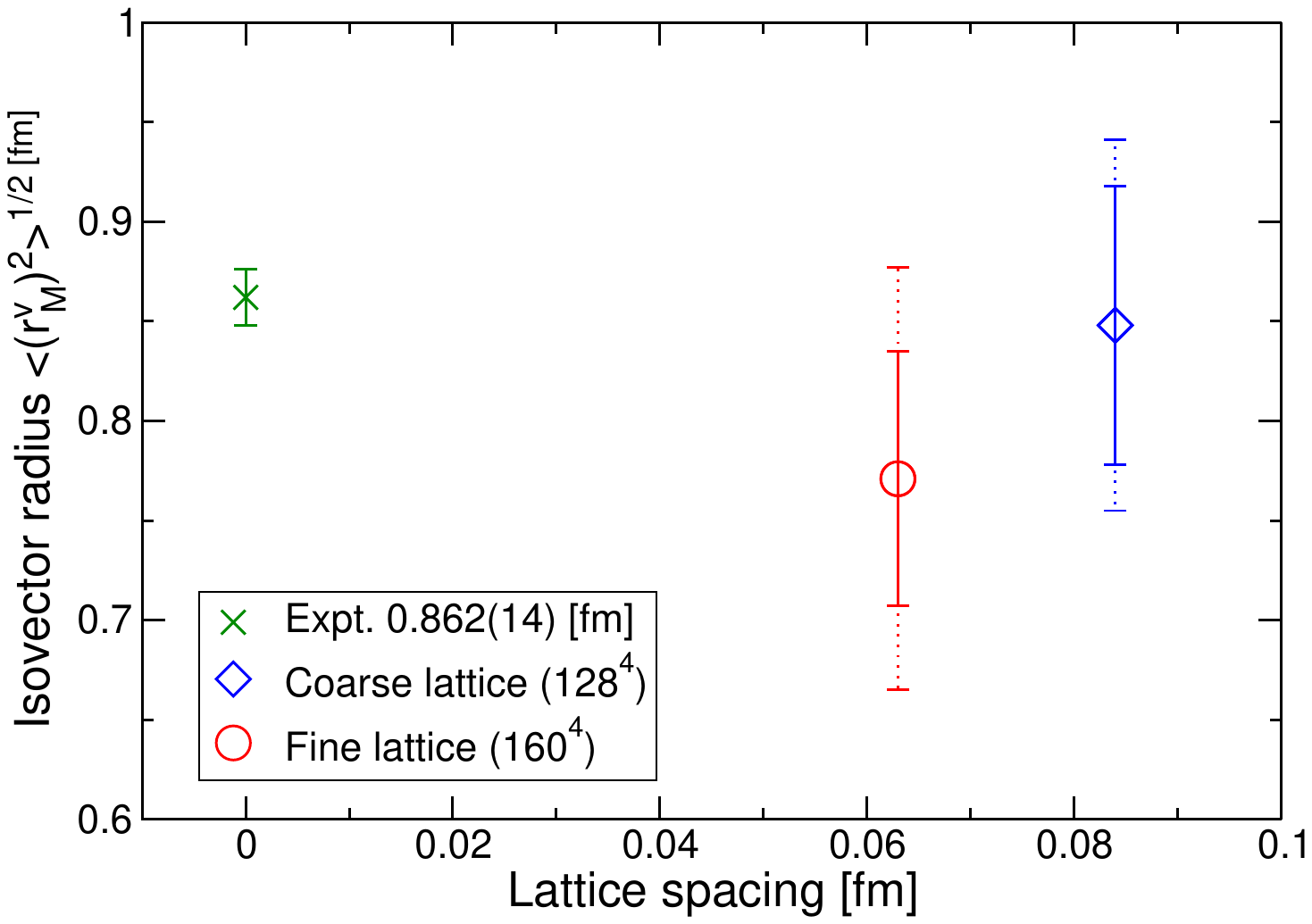}
\includegraphics[width=0.32\textwidth,bb=0 0 792 612,clip]{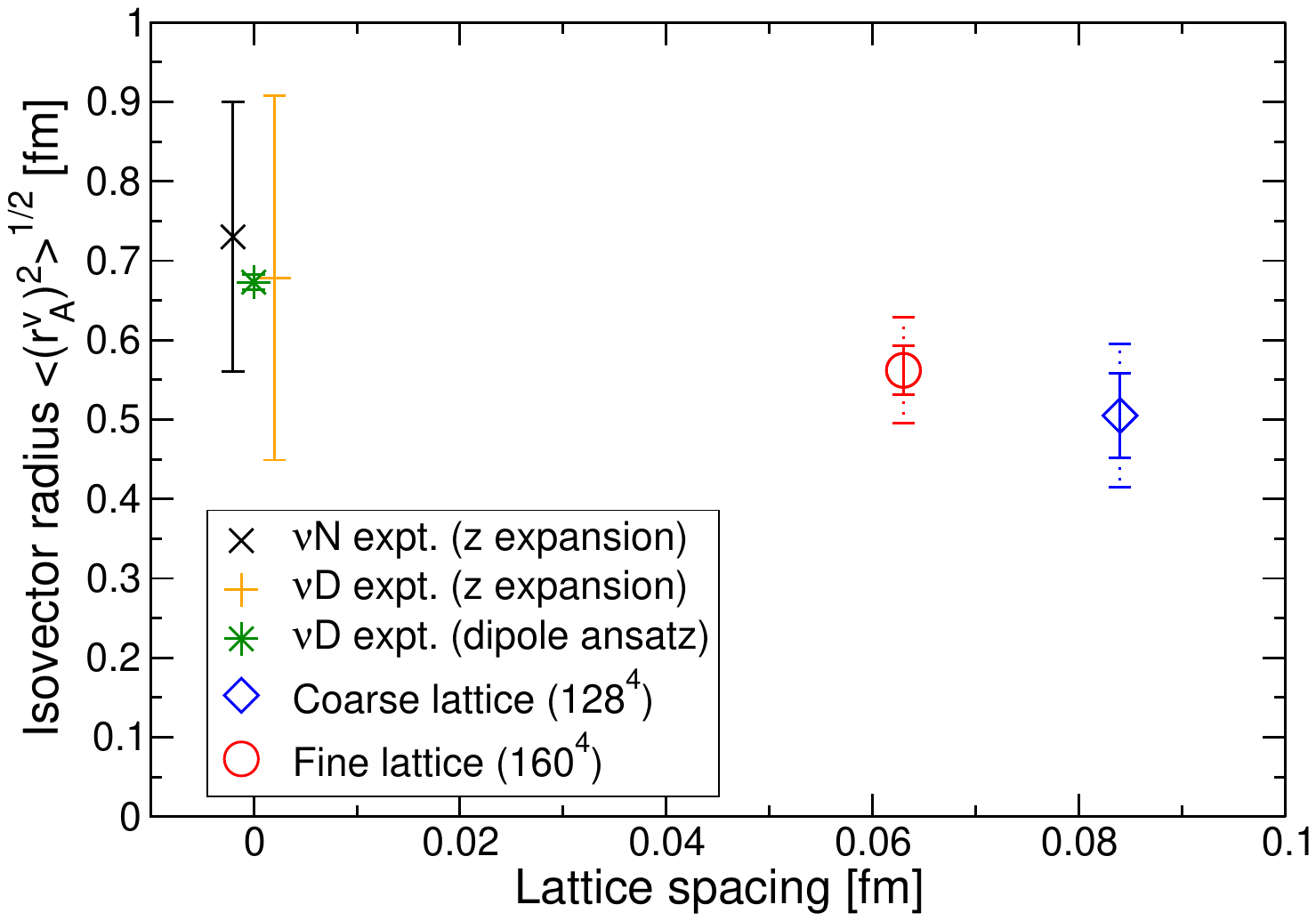}
\caption{Summary for our best estimates and the experimental values 
forthe axial-vector coupling (top, left), isovector magnetic moment (top, right) and three kinds of the isovector RMS radius: electric (bottom, left), magnetic (bottom, center) and axial (bottom, right). The inner error bars show the statistical error,
while the outer error bars evaluated by both the statistical and systematic errors added in quadrature. Uncertainties from the excited-state contamination and the violation of the dispersion relation are taken into account as the systematic errors.
This figure is reprinted from Ref.~\cite{Tsuji:2023llh}.}
\label{fig:continuum_limit}
\end{figure*}

Figure \ref{fig:continuum_limit} shows 
the lattice spacing $a$-dependence for these five quantities.
The inner error bars represent the statistical uncertainties, while the outer error bars represent
the total uncertainties given by adding the statistical
errors and systematic errors in quadrature. The systematic errors take into account
uncertainties stemming from the excited-state contamination and so on
(see details \textbf{}~Ref.~\cite{Tsuji:2023llh}).

%the lattice discretization effect {\color{red}estimated through} the dispersion relation.

Let us first discuss the size of the finite lattice spacing effect on the axial-vector coupling $g_A$, 
that is precisely measured by the experiments. 
The axial-vector coupling $g_A=F_A(q^2=0)$ is directly determined from the ratio~(\ref{eq:fa_def})
at zero momentum transfer without the
$q^2$-extrapolation.
%to the zero momentum point.
In the top-left panel of Fig.~\ref{fig:continuum_limit}, 
the two results obtained at different lattice spacings can reproduce the experimental value within statistical precision of at most 2\%.
This implies that the discretization error on the axial-vector coupling is less than 2\%, which is well controlled in our calculations.
In the top-right panel of Fig.~\ref{fig:continuum_limit},
the small discretization error, which is less than 1\%,  is also observed for the magnetic moment $\mu_v$ evaluated by Eq.~(\ref{eq:gm_def}),
the two results for the magnetic moment are both 5-6\% smaller than the experimental value.
However, recall that
the magnetic moment is not accessible without the 
$q^2$-extrapolation to the zero momentum point,
and hence 
more comprehensive investigations with
the derivative of form factor (DFF) method~\cite{Ishikawa:2021eut}.
are necessary in order to fully resolve the current discrepancy.

Apart from the question of whether the results are consistent with the experimental values, both quantities, $g_A$ and $\mu_v$, do not seem to be subject to large discretization errors. 
In contrast, the RMS radii,
which are determined from the form-factor slope at the zero momentum point,
may suffer from the $O(qa)$ discretization effects that do not appear in $g_A$ and $\mu_v$.
Indeed, as shown in three bottom panels of 
Fig.~\ref{fig:continuum_limit}, the presence of the discretization errors is clearly visible for the isovector RMS radii.
Their sizes can be estimated as about 10\%, regardless of the channel.

Especially, $\sqrt{\langle (r^v_E)^2 \rangle}$
can be evaluated with a statistical error of less than 5\% accuracy, while the magnitude of the discretization uncertainty is much larger than the statistical one.
Therefore, as shown in the bottom-left panel of Fig.~\ref{fig:continuum_limit}, the certain difference between the two results is clearly observed in $\sqrt{\langle (r^v_E)^2 \rangle}$, which is unexpectedly large. 
However, this observation may bridge the gap between our lattice results and experimental values.
Similarly, the finite lattice spacing effect observed in $\sqrt{\langle (r^v_A)^2 \rangle}$ as shown in
the bottom-right panel of Fig.~\ref{fig:continuum_limit}
tend to fill the difference between lattice QCD results and experimental values.
It is important to emphasize here that the total errors in the axial radius obtained at two lattice spacings are much smaller than the two estimations obtained from the model-independent $z$-expansion analysis for both $\nu N$ and $\nu D$ scattering data.

%
%  FIG.55
%

\section{Summary}
\label{sec:summary}

We have calculated 
%the nucleon axial-vector coupling, and 
the nucleon form factors
in the vector and axial-vector channels
using the second PACS10 ensemble generated at the physical point on a $(10\;{\mathrm{fm}})^4$ volume.
The PACS10 gauge configurations are generated by the PACS Collaboration with the stout-smeared $O(a)$ improved Wilson quark action and Iwasaki 
gauge action.

Since the continuum-limit extrapolation requires results from at least {\it three lattice spacings}, we have investigated the systematic uncertainties associated with the finite lattice spacing on
$g_A$ and isovector RMS radii
from the difference between the current results obtained at two lattice spacings.
It was found that the the finite lattice spacing effect on $g_A$ is kept below the statistical error of less than 2\%, which is currently achieved in our calculations, while both results
of $g_A$ obtained at two lattice spacings
reproduce the experimental value within their statistical precisions.  
Therefore, the lattice discretization 
effect on $g_A$ is negligibly small in our calculations.
On the other hand, 
the systematic errors associated
with the finite lattice spacing
on the isovector RMS radii are
about 10\%
and cannot be ignored regardless of channel.

Needless to say that additional lattice simulations using {\it the third PACS10 ensemble} is required for achieving a comprehensive study of the discretization uncertainties and then taking the continuum limit of our target quantities.
Such planning is now underway.

\section*{Acknowledgement}
We would like to thank members of the PACS collaboration for useful discussions.
R.~T. is supported by the RIKEN Junior Research Associate Program.
R.~T. acknowledge the support from Graduate Program on Physics for the Universe (GP-PU) of Tohoku University.
K.-I.~I. is supported in part by MEXT as ``Feasibility studies for the next-generation computing infrastructure".
%K.~S. is supported by JST, The Establishment of University Fellowships towards the creation of Science Technology Innovation, Grant Number JPMJFS2106.
%We also thank Y. Namekawa for his careful reading of the manuscript.
Numerical calculations in this work were performed on Oakforest-PACS in Joint Center for Advanced High Performance Computing (JCAHPC) and Cygnus  and Pegasus in Center for Computational Sciences at University of Tsukuba under Multidisciplinary Cooperative Research Program of Center for Computational Sciences, University of Tsukuba, and Wisteria/BDEC-01 in the Information Technology Center, the University of Tokyo. 
This research also used computational resources of the K computer (Project ID: hp1810126) and the Supercomputer Fugaku (Project ID: hp20018, hp210088, hp230007) provided by RIKEN Center for Computational Science (R-CCS), as well as Oakforest-PACS (Project ID: hp170022, hp180051, hp180072, hp190025, hp190081, hp200062),  Wisteria/BDEC-01 Odyssey (Project ID: hp220050) provided by the Information Technology Center of the University of Tokyo / JCAHPC.
The  calculation employed OpenQCD system(http://luscher.web.cern.ch/luscher/openQCD/). 
This work is supported by the JLDG constructed over the SINET5 of NII,
This work was also supported in part by Grants-in-Aid for Scientific Research from the Ministry of Education, Culture, Sports, Science and Technology (Nos. 18K03605, 19H01892, 22K03612, 23H01195, 23K03428) and MEXT as ``Program for Promoting Researches on the Supercomputer Fugaku'' (Search for physics beyond the standard model using large-scale lattice QCD simulation and development of AI technology toward next-generation lattice QCD; Grant Number JPMXP1020230409).

%--- bibliography ---------------------------------------------------  
\bibliographystyle{man-apsrev}
\bibliography{skelton}

\end{document}